# *Epsilon-Near-Zero (ENZ) Metamaterials and Electromagnetic Sources: Tailoring the Radiation Phase Pattern*

*Andrea Alù[(1)], Mário G. Silveirinha[(1,2)], Alessandro Salandrino[(1)], and Nader Engheta[(1),*]*

*(1) University of Pennsylvania, Department of Electrical and Systems Engineering, Philadelphia, Pennsylvania 19104, U.S.A.*

*(2) Universidade de Coimbra, Department of Electrical Engineering– Instituto de Telecomunicações, Portugal*

*Abstract*

In this work, we investigate the response of epsilon-near-zero (ENZ) metamaterials and plasmonic materials to electromagnetic source excitation. The use of these media for tailoring the phase of radiation pattern of arbitrary sources is proposed and analyzed numerically and analytically for some canonical geometries. In particular, the possibility of employing planar layers, cylindrical shells or other more complex shapes made of such materials in order to isolate two regions of space and to tailor the phase pattern in one region, fairly independent of the excitation shape present in the other, is demonstrated with theoretical arguments and some numerical examples. Physical insights into the phenomenon are also presented and discussed together with potential applications of the phenomenon.

## 1. Introduction

The growth of interest in metamaterials and plasmonic materials has recently led not only to novel and interesting possibilities for microwave, infrared and optical applications, but also to several conceptual advancements in the fundamentals of the electromagnetic theory. Limits that were considered insurmountable in conventional setups have indeed been shown to be possibly surpassed when special materials are employed. Examples of these achievements, such as sub-diffraction transport of information and sub-wavelength focusing, have been proven theoretically and experimentally in the recent technical literature.

In particular, materials with anomalous values of their effective permittivities and/or permeabilities have been analyzed in detail, owing to their anomalous and often counterintuitive wave interaction. For instance, a slab made of a material with simultaneously negative permittivity and permeability at the desired frequency of operation has been shown to potentially focus the sub-wavelength details of an object as a "superlens" [1]. Such materials, also named double-negative (DNG) [2], have been realized at microwave frequencies by embedding properly designed resonant electrical and magnetic inclusions in a periodic lattice [3] and several attempts for extending these concepts to higher frequencies (up to the visible domain) are being recently conducted by several groups [4]-[7].

Materials with negative permittivity (ε-negative, ENG) at these higher frequencies, i.e., infrared and visible, are already available in nature, even with relatively low losses, and they are represented by noble metals and polar dielectrics. Generally the permittivities of such materials follow Drude or Drude-Lorenz dispersion models [8]-[9], that describe the frequency variation of the resonances of their molecular components, responsible for the anomalous values of permittivities typical of the specific range of frequencies where these resonances take place. Their plasma frequency $f_p$, which is the frequency at which the real part of their permittivity effectively goes to zero, usually lies in the terahertz regime for polar dielectrics and some semiconductors [10] and in the visible and ultra-violet for noble metals [8]-[11]. The interest in such plasmonic materials is quite relevant nowadays, since the bulk and surface plasmonic resonances characterizing particles made of these materials may be exploited in several ways. The advantage of utilizing such media clearly resides in the fact that they are already available in nature, overcoming the difficulties of manufacturing such materials for higher frequencies.

The window of frequency in which the permittivity is low, i.e., near the plasma frequency, has also become a topic of research interest in several potential applications. The first attempts to build a material with low permittivity at microwave frequencies date back to several decades ago, where their use was proposed in antenna applications for enhancing the radiation directivity [12]. Similar attempts have been presented over the years with analogous purposes [13]-[16]. Other more recent investigations of the properties of ε-near-zero (ENZ) materials and metamaterials and their intriguing wave interaction properties have been reported in [17]-[18]. In particular, Ref. [18] addresses and studies the possibility of designing a bulk material, impedance matched with free space, but whose permittivity and permeability are simultaneously very close to zero.

In all these works the main attempt has been to exploit the low wave number (index near zero) propagation in such materials, which might provide a relatively small phase variation over a physically long distance in these media. When interfaced with materials with larger wave number, this implies the presence of a region of space with almost uniform phase distribution, providing the possibility for directive radiation towards the broadside to a planar interface, as proposed over the years for antenna applications. Also, in [18] the possibility of utilizing the matched low-index metamaterial for transforming curved phase fronts into planar ones has been suggested, exploiting the matching between the aforementioned metamaterial and free space.

Our group has recently proposed several different potential applications of ENZ and/or μ-near-zero (MNZ) materials for different purposes. Relying on the directivity enhancement that such materials may provide, we have shown how it is possible to cover an opaque screen with low-index materials in order to dramatically enhance the transmission through a sub-wavelength narrow aperture in the screen [19]. This effect can be explained in terms of the leaky-waves supported by such grounded layers, which constitute directive leaky-wave antennas [20]. Relying on a different mechanism, we have recently shown the possibility of squeezing electromagnetic energy through plasmonic ENZ sub-wavelength narrow channels, and demonstrated analytically and numerically how this effect may improve the transmission at a sharp waveguide bend [21].

In a different context, in [22] we have shown how ENZ materials may be used as covers to cancel the scattering from dielectric or even conducting objects, drastically reducing their total scattering cross sections and making the covered objects practically undetectable to an external observer. This effect is related to the negative polarizability that such covers may exhibit, due to their low permittivity with respect to the background medium.

In this paper, motivated by these exciting anomalous properties and potential applications of ENZ materials, we investigate in detail their behavior in the presence of electromagnetic sources. In particular, we investigate the possibility of manipulating the phase fronts of such sources for obtaining anomalous imaging, lensing and radiative effects. This is done for planar and cylindrical geometries in analytical terms and for other more complex shapes using numerical full-wave simulations. These results may provide interesting new possibilities for imaging and radiative tools at infrared and optical frequencies, where such plasmonic ENZ materials may be readily available in nature. Similar results may be obtained at lower frequencies employing engineered metamaterials.

## 2. *Heuristic Analysis*

In the limit in which the permittivity $\varepsilon$ of a given material is zero at a specific frequency, Maxwell's equations under the $e^{-i\omega t}$ convention in a source-free region are written in the form:

$$\begin{aligned}\nabla \times \mathbf{H} &= \mathbf{0} \\ \nabla \times \mathbf{E} &= i\omega\mu_0 \mathbf{H}\end{aligned} \quad (1)$$

with $\mu_0$ being the free-space permeability, that is assumed to be also the permeability of the zero-permittivity material. Eq. (1) implies that the magnetic field is a curl-free vector and that wave propagation in this material can happen only with phase velocity being infinitely large (satisfying the "static-like" equation $\nabla^2 \mathbf{E} = \mathbf{0}$, which is obtained directly from (1)).

The fact that the phase variation in such material may be effectively very small compared to the free-space wavelength results in some interesting consequences. Imagine to have a system like the one shown in Fig. 1, in which a given phase front impinges on the entrance side of an object (side A in the figure) made of such an ENZ material. If somehow we manage to couple electromagnetic energy into this material (which is not trivial by itself, since the impedance mismatch would be present), the phase front at the exit side (side B) should conform to the shape of the exit face, since there is essentially no phase variation in the wave propagation inside the material. This implies that in principle with the use of such plasmonic (ENZ) materials we can manipulate a given impinging phase front and transform its phase distribution into a desired shape by properly tailoring the exit side of the ENZ slab. This of course may have important implications in imaging and communications technology, since we may speculate that by designing the conformal face of the exit side B one can shape the phase front to a given desired pattern, independent on the form of the impinging wave.

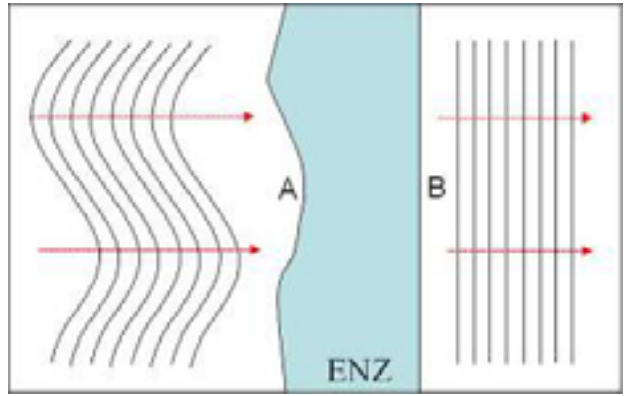

Figure 1 – (Color online) A curved phase front impinges on an ENZ material in its entrance side (side A). At the exit side (side B) the phase front is conformal to the exit surface, in this case a planar phase font, due to the small phase variation inside the material.

In this paper, we present analytical solutions and numerical simulations for certain canonical geometries involving ENZ materials in proximity of sources, which may reveal the conditions and limitations of the previous heuristic prediction in realistic setups.

3.  *Planar ENZ slab*

Consider as a first case the geometry of Fig. 2, which shows an infinitely extent planar slab of ENZ material, with permittivity $\varepsilon_s$, permeability $\mu_0$ and thickness $d_{slab}$, excited by a TM plane wave with magnetic field $\mathbf{H}(\beta) = H_0 \hat{\mathbf{y}} e^{i\sqrt{k_0^2 - \beta^2}(z + d_{source})} e^{i\beta x}$. Here, $k_0 = \omega\sqrt{\varepsilon_0 \mu_0}$ represents the wave number in free space, $H_0$ is a generic complex amplitude, $d_{source}$ represents a given reference plane and $\beta = k_0 \sin\theta_i$, where $\theta_i$ is the angle of incidence as indicated in the figure. It is straightforward to calculate transmission and reflection coefficients for this simple geometry. Defining $R(\beta)$ as the complex reflection coefficient for the magnetic field at the entrance face of the slab (at $z = 0$) and $T(\beta)$ as the complex transmission coefficient at the exit face (at $z = d_{slab}$), interesting limiting expressions for this problem may be derived by taking the limit for the slab's permittivity going to zero. The transmission and reflection coefficients yield the following expressions in this limit:

$$\lim_{\varepsilon \to 0} R(\beta) = \begin{cases} -\dfrac{k_0 d_{slab} e^{i k_0 d_{source}}}{2i + k_0 d_{slab}} & \beta = 0 \\ -e^{-j k_0 d_{source}} & \beta \neq 0 \end{cases}$$

$$\lim_{\varepsilon \to 0} T(\beta) = \begin{cases} \dfrac{e^{i k_0 d_{source}}}{1 - i\dfrac{k_0 d_{slab}}{2}} & \beta = 0 \\ 0 & \beta \neq 0 \end{cases} \qquad (2)$$

Eqs. (2) imply that no transmission through such a zero-permittivity slab occurs, unless the impinging plane wave comes exactly at broadside ($\theta_i = 0$), for which the transmission is non-zero and is inversely proportional to the electrical size of the slab. It seems evident here how such an ENZ slab would act as an ideal angular filter in the limit of $\varepsilon_s = 0$, with an "anomalous" discontinuity in the transmission coefficient, as one moves the angle from broadside incidence to any other, even infinitesimally small, angle. The magnetic field distribution inside the slab would be identically zero unless $\beta = 0$, for which particular case $\mathbf{H}(0)$ would be constant inside the slab (to fulfill the curl-free condition in (1)) with value $\mathbf{H}(0) = H_0 \dfrac{e^{i k_0 d_{source}}}{1 - i\dfrac{k_0 d_{slab}}{2}} \hat{\mathbf{y}}$. The electric field would decay exponentially inside the slab for $\beta \neq 0$, whereas for normal incidence, the electric field, being directed along $\hat{\mathbf{x}}$, would be linearly varying along $z$ to satisfy (1) and to match the continuity conditions at the two boundaries.

It is worth underlining how for any $\beta \neq 0$ the $\varepsilon_s = 0$ slab acts as a perfect magnetic boundary in this geometry, yielding a 180° phase shift for reflection of the magnetic field at the entrance face. The anomalous behavior with partial tunneling of the field at $\beta = 0$ is explained with a polaritonic resonance of such a slab, as we discuss in the following.

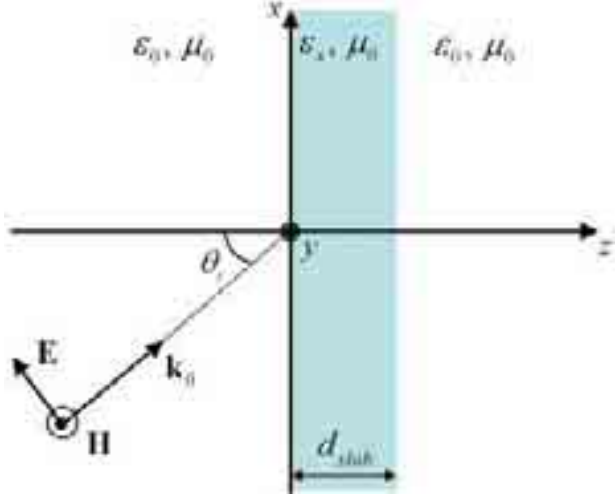

Figure 2 – (Color online) A planar slab of permittivity $\varepsilon_s$ excited by a TM plane wave in a suitable Cartesian coordinate system.

Fixing the value of the slab permittivity $\varepsilon_s$ to a non-zero, but low value, we can study the behavior of the transmission function when the angle of incidence of the impinging plane wave is varied, in order to understand how the limiting discontinuous response predicted by (2) is reached when $\varepsilon_s$ becomes identically zero. In this way, we can predict the realistic response of an ENZ planar slab near its plasma frequency to a source excitation, satisfying all the physical constraints of continuity and finiteness of the fields. Also some physical insights may be gained by this analysis, as we present below.

The transmission coefficient for such a simple problem can be written in compact form as:

$$T(\beta) = \frac{e^{ik_{t0}d_{source}}}{\cos(k_{ts}d_{slab}) + i\frac{\left[\beta^2\left(\varepsilon_0^2 + \varepsilon_s^2\right) - k_0^2\varepsilon_s\left(\varepsilon_0 + \varepsilon_s\right)\right]}{2\varepsilon_0\varepsilon_s k_{t0}k_{ts}}\sin(k_{ts}d_{slab})}, \qquad (3)$$

where $k_{t0} = \sqrt{k_0^2 - \beta^2}$ and $k_{ts} = \sqrt{k_{slab}^2 - \beta^2}$. The sign of the square root for $k_{t0}$ should be chosen to satisfy the radiation condition, i.e., its imaginary part should be non-negative, whereas the branch choice for $k_{ts}$ does not influence the solution of (3).

Eq. (3) clearly shows that the Brewster angle for this problem, which corresponds to the polariton resonance of the structure under analysis [19], [23]-[25], is given by the simple relation $\sin(k_{ts}d_{slab}) = 0$ for which $|T| = 1$. Since we are not considering electrically thick slabs, and the wave number in the ENZ slab is small, the only available polariton resonance, for which the wave tunnels completely through the slab despite the huge mismatch between free space and the ENZ material, is represented by the condition $\beta_{pol} = k_{slab} \simeq 0$.

The "quality factor" $Q$ for this resonance in terms of the "angular fractional bandwidth" can be found by considering the complex root for $\beta = \beta_r + i\beta_i$ of the denominator in (3) and is given by $Q_\beta = \beta_r / (2\beta_i)$. Interestingly, its value for ENZ materials does not depend on $\varepsilon_s \ll \varepsilon_0$ and is given by the closed form expression:

$$Q_\beta = \frac{k_0 d_{slab}}{2}. \qquad (4)$$

Since this expression is proportional to the inverse of the fractional angular bandwidth of the system, this shows the two interesting aspects: (a) for a fixed material parameter $\varepsilon_s \ll \varepsilon_0$ the peak in the transmission coefficient at $\beta_{pol}$ narrows linearly its angular bandwidth with the thickness of the slab, and (b) for a fixed thickness of the slab the interval $\Delta\beta$ of angles that actually tunnel through the slab narrows down when $\varepsilon_s \to 0$. Since Eq. (4) ensures that the fractional angular bandwidth $Q_\beta^{-1} = \Delta\beta / \beta_{pol}$ remains constant with a variation of $\varepsilon_s$ (in the ENZ limit) and $\beta_{pol} = k_{slab} = \omega\sqrt{\varepsilon_s \mu_0}$ is proportional to $\sqrt{\varepsilon_s}$, the angular bandwidth of transmission $\Delta\beta$ reduces proportionally with $\sqrt{\varepsilon_s}$.

These results are consistent with the limiting case of Eq. (2), for which $\varepsilon_s = 0$ and the interval of angles that tunnels through the slab becomes infinitesimally narrow, i.e., $\Delta\beta = 0$ (since $\beta_{pol} = 0$ as well, $Q_\beta$ may remain finite in this limit and it may still satisfy (4)). Following these arguments, a closed form expression for the interval of phase vectors (centered around $\beta_{pol} = k_{slab}$) that tunnels with sufficient transmission (half of the impinging power) through such a slab is given by:

$$\Delta\beta = \frac{2k_{slab}}{k_0 d_{slab}}. \qquad (5)$$

In terms of angles, considering that $\Delta\beta \ll k_0$ and that $\beta_{pol} \simeq 0$, Eq. (5) can be written as:

$$\Delta\theta \simeq \frac{\Delta\beta}{k_0} = \frac{2\sqrt{\varepsilon_s/\varepsilon_0}}{k_0 d_{slab}}, \qquad (6)$$

which shows how the transmitted angular band narrows down when the relative permittivity of the slab approaches zero, or when the slab thickness is increased.

Figure 3 shows the transmission coefficient for different values of the slab thickness $d_{slab}$, and for two different $\varepsilon_s$ close to zero, consistently with formula (5). In particular, you may notice how there is always a peak of total transmission, corresponding to the case $\beta = \beta_{pol}$, with the angular bandwidth $\Delta\beta$ decreasing with an increase in $d_{slab}$ or a decrease in $\varepsilon_s$. In the figure it is clearly shown how the position of the polariton resonance is independent of the thickness of the slab, since this anomalous transmission does not rely on a resonance of the slab (as it would be for thicker slabs where $k_s d_{slab}$ is a multiple of $\pi$, as in any Fabry-Perot resonator), but rather it is a *material* resonance due to the fact that at this specific angle the longitudinal component of the wave number in the ENZ material is null. We note from these figures how the required value of slab permittivity $\varepsilon_s$ has to be very close to zero in order to obtain a strong angular selectivity, and therefore a strong modification of the phase pattern at the exit face of the slab when a collection of plane waves impinges on the slab, since according to Eq. (6) the angular bandwidth has a slow variation with permittivity, i.e., it is proportional to $\sqrt{\varepsilon_s}$. When $\varepsilon_s \to 0^+$ the peaks in transmission slowly move towards broadside with a simultaneous decrease in the angular bandwidth of transmission. Exactly at broadside the transmission follows the decrease with the slab thickness $d_{slab}$ predicted by Eq. (2) and confirmed by Fig. 3. In the limit of $\varepsilon_s = 0^+$ the transmission peak vanishes, and the slab does not support a real polariton resonance, since its angular bandwidth becomes identically zero, even though its response at broadside is maintained, as Eq. (2) testifies, confirming partial wave tunneling at this specific angle.

For negative low values of $\varepsilon_s$, again the polariton resonance is below cut-off, since $k_{slab}$ is not a real quantity. The peak of transmission clearly remains at broadside, even though it is less than unity and decreases exponentially with the slab thickness. For reasonably small $|k_s d_{slab}|$ also $\varepsilon_s = 0^-$ would act as an angular filter with similar properties of modification of the impinging phase pattern to conform to the exit face of the slab.

The presence of material losses can noticeably affect the above results, since the polariton resonance may be drastically weakened by the presence of realistic losses in the slab material. In this sense, a larger angular bandwidth implies a higher robustness to the material loss, as in any resonant system.

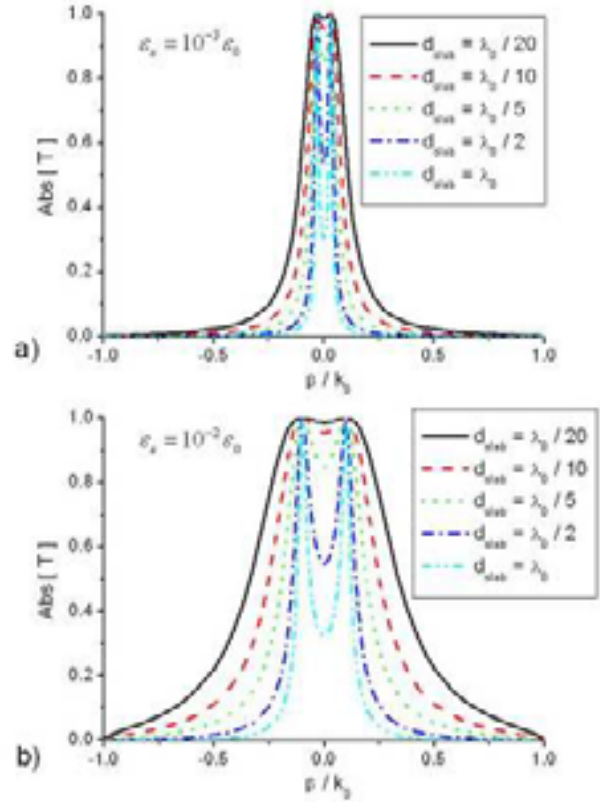

Figure 3 – (Color online) Transmission coefficient for the geometry of Fig. 2 with different slab thickness $d_{slab}$ and permittivity $\varepsilon_s$ near zero.

The TE polarization does not support polariton resonances in this setup, as it can be easily noticed by applying duality, since the slab has the same permeability as the background medium. The peak of transmission is expected at broadside, which coincides with the TM analysis, with a smooth degradation of transmission with an increase in the angle of incidence, and converges towards zero for $\theta_i = \pi/2$. No high angular selectivity is observed in this TE polarization, due to the lack of polariton resonances.

To summarize, the previous analysis shows that an ENZ planar slab may act as an angular filter for the TM polarization, allowing transmission only for a specific narrow angular width close to the normal. Consistently with the heuristic prediction that the phase at the exit face should be constant all over the surface, the plane waves that tunnel through such a system have always a propagation direction very close to the normal. The previous rigorous analysis has shown that a polariton resonance is responsible for the wave tunneling, which actually happens at $\beta_{pol} = k_{slab}$. One may further notice how the ENZ slab effectively isolates the entrance side from the exit side. For the angles that do not tunnel through the slab, the slab acts as a perfect magnetic conductor, isolating electrically the two faces. The only allowed response between the two sides is related to the excitation of the material polaritons supported by the structure.

Let us suppose now that a more complex wave front impinges on the entrance face of the slab, composed of a wide spectrum of plane waves. In this case, it is clear from the previous study that only the components of the spectrum with wave number close to $\beta_{pol} = k_{slab} \simeq 0$ and TM polarization would tunnel through the system, and, as predicted heuristically in Fig. 1 and shown in the previous lines, the phase front can be modified by the slab in order to resemble the planar interface at the exit side of the slab. All other wave components would be reflected back by the system.

Suppose for instance that an infinitesimal 2-D magnetic source, like a magnetic current line, is positioned in front of the slab. For simplicity of analysis we assume that at a distance $d_{source} = \lambda_0$ from the slab the magnetic field has a point-like distribution, i.e., the impinging magnetic field has the distribution $\mathbf{H}_{imp} = H_0 \hat{\mathbf{y}} \delta(z + d_{source}) \delta(x)$. In the free-space region, the magnetic field distribution can be written as an integral superposition of plane waves with magnetic field $\mathbf{H}(\beta)$ as defined above, each of which is transmitted with amplitude $T(\beta)$ through the slab.

Figure 4 depicts the amplitude and phase distributions under such an excitation at the exit face of an ENZ slab with $\varepsilon_s = 10^{-4} \varepsilon_0$, while different thicknesses $d_{slab}$ are considered. The results were calculated by analytically solving the problem as superposition of plane waves with proper weighting coefficients given by the transmission coefficients $T(\beta)$ given by (3). All the curves are normalized to their relative maximum peak. The dotted thin lines in the plots refer to the case in which the slab is removed, and the field is sampled at the coordinate $z = d_{noslab} = \lambda_0$. One can notice

-12-

how the presence of the slab implies a very smooth variation of the phase along several wavelengths, with a drastic modification of the radiation pattern, due to the total reflection of most part of the impinging spectrum of plane waves. This corresponds to a drastic modification of the radiation pattern at the exit side of the slab. Also the amplitude is more uniform along the exit side, when compared to the case in which the slab is removed. These results confirm the heuristic predictions that an ENZ material may be employed to modify effectively the imaging and radiation pattern of a given source, which may be useful in different applications.

Figure 5, as an example, shows the power flow for a line current placed at distance $d_{source} = 2\lambda_0$ from the entrance side of a slab with $\varepsilon_s = 10^{-4}\varepsilon_0$ and $d_{slab} = \lambda_0/5$. It can be seen how the power flow is significantly redirected by the polariton resonance present inside the slab and that the electromagnetic field becomes very close to a plane wave distribution directed towards broadside at the exit side.

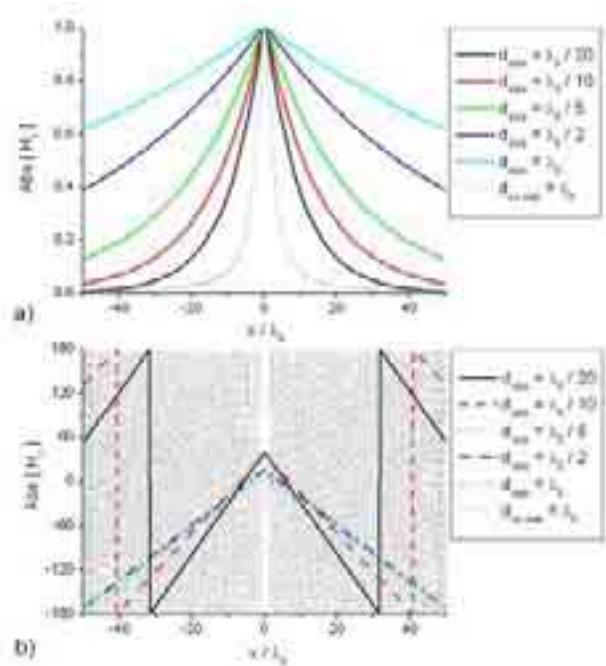

Figure 4 – (Color online) Normalized amplitude (a) and phase (b) distributions at the exit face for an infinite magnetic line current placed at distance $d_{source} = \lambda_0$ from the entrance side of an ENZ slab with $\varepsilon_s = 10^{-4}\varepsilon_0$ for different thicknesses $d_{slab}$.

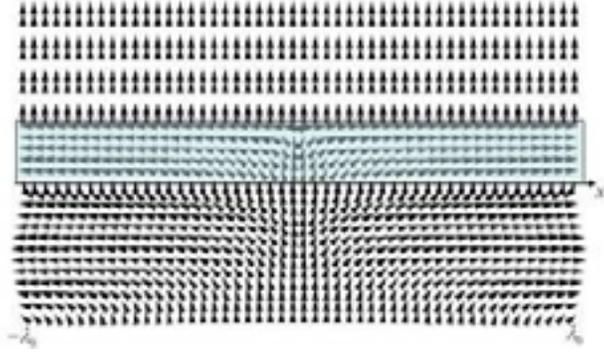

Figure 5 – (Color online) Power flow (real part of the Poynting vector) for a source placed at distance $d_{source} = 2\lambda_0$ from the entrance side of a slab with $\varepsilon_s = 10^{-4}\varepsilon_0$ and $d_{slab} = \lambda_0/5$.

We have also considered the case of bending the exit side of the slab in order to verify whether the phase pattern follows the same perturbation of the shape of the exit face. Figure 6 shows the magnetic field distribution induced by an electrically short electric dipole placed in front of an ENZ slab, with the exit face bent in order to have a finite curvature. The full-wave simulation has been performed with finite-integration-technique commercial software (CST Microwave Studio$^{TM}$ [26]) considering a realistic setup, i.e., a finite slab and a dispersive Drude-like lossy material with:

$$\varepsilon_s = \left[1 - \frac{f_p^2}{f(f+i\gamma)}\right]\varepsilon_0, \quad (7)$$

where $f$ is the operating frequency, $f_p$ is the plasma frequency at which $\text{Re}[\varepsilon_s] \simeq 0$ and $\gamma$ is the damping factor taking into account material losses. The geometry and the material properties are displayed in the figure and in the caption. The excitation is represented by a short electric dipole lying horizontal and parallel to the entrance face in the figure. We notice how the exit phase distribution is conformal to the exit face shape, as we may expect following the same reasons discussed in this paragraph. This shows how the ENZ material allows, under proper conditions, modification of the phase distribution of an impinging wave, by properly tailoring the shape of the bulk material.

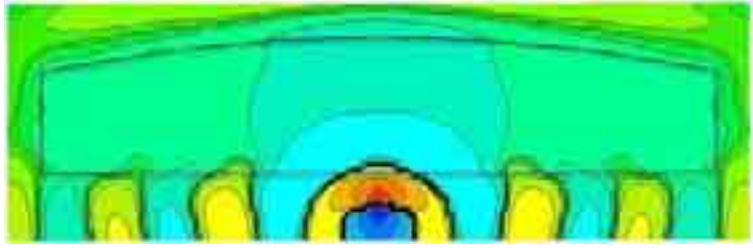

Figure 6 – (Color online) Snapshot in time of the magnetic field distribution induced by an electric small dipole placed in front of an ENZ slab with a bent exit face. Here the simulation has considered a realistic 3-D setup: the entrance face is a planar square of 3 x 3 wavelengths at the operating frequency $f_0$; a small electric dipole, parallel to the entrance face, is placed in its close proximity. The ENZ material composing the slab follows the material dispersion [11] with $f_p = f_0$ and $\gamma = 10^{-2} f_0$, representing material losses for a metamaterial or a plasmonic material. Brighter (more red) colors correspond to higher amplitude of the instantaneous local magnetic field.

We have verified a similar effect of phase front reshaping in a different setup by use of finite-element-method commercial software (COMSOL Multiphysics™ [27]) for the concave-lens-like 2-D structure depicted in Figure 7. The figure shows how the structure may convert planar phase fronts, impinging on the bottom of the structure, into converging fronts that follow the curvature of the exit face. Here the slab is assumed to be made of silver, including realistic losses in the model, with permittivity $\varepsilon_s = 0.016 + i0.42$, which is the measured permittivity of $Ag$ for $\lambda_0 = 325\,nm$ [11]. We will report similar effects and others obtained with more complex geometries in the last section of the manuscript using a moment-method full-wave solution.

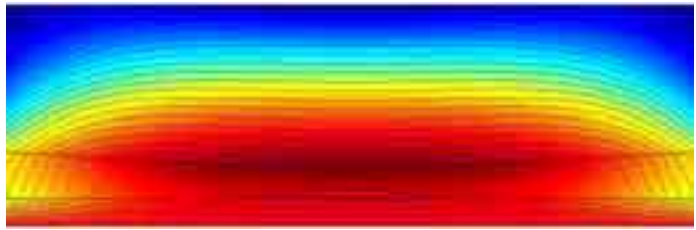

Figure 7 – (Color online) Magnetic field distribution induced by a plane wave incident on a 2D ENZ slab with concave exit face. The slab width is $1.2\lambda_0$ and it is assumed to be made of realistic silver with permittivity $\varepsilon_s = 0.016 + i0.42$ at the frequency for which $\lambda_0 = 325\,nm$, following experimental data [11]. Brighter (more red) colors correspond to higher amplitude values of the magnetic field. The dark lines represent equiphase fronts.

## 4. Cylindrical ENZ shell

The cylindrical geometry may provide further verification of the possibility of modifying a given phase pattern by employing ENZ materials. Consider for instance Fig. 8, i.e., a cylindrical ENZ shell of permittivity $\varepsilon_s \ll \varepsilon_0$ surrounding a hollow cavity. Again all the permeabilities are considered the same as the one of free space. Applying the same concepts as in the previous paragraph, we intuitively expect that it may be possible to excite the structure with an arbitrary phase pattern in the outside region and obtain an angularly uniform phase distribution in the hollow cavity. Again, the ENZ shell in this configuration may act as an ideal isolator between the two regions of space, producing in the hollow core the desired phase distribution, weakly dependent on the source distribution on the other side of the shell. Further than confirming the previous heuristic prediction in a curved geometry, this effect may have interesting implications for isolating a closed region of space from the phase variation that impinges from outside.

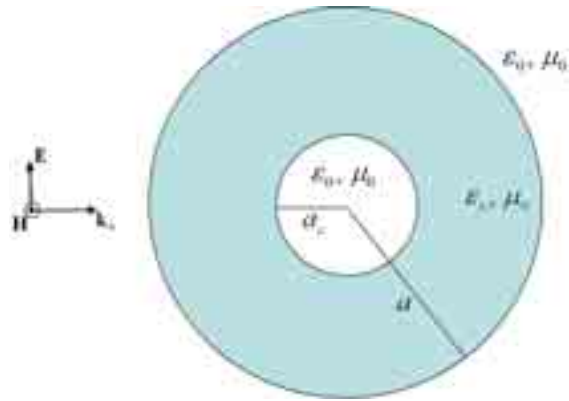

Figure 8 – (Color online) A cylindrical shell with permittivity $\varepsilon_s$ much smaller than $\varepsilon_0$, and radius $a$ enclosing a hollow cavity of radius $a_c$.

If a TM plane wave impinges on this structure, as depicted in the figure, the problem can be solved using the formal Mie approach. In particular, expanding the impinging plane wave in terms of Bessel functions, i.e., $\mathbf{H}_i = H_0 \hat{\mathbf{z}} \sum_{n=-\infty}^{\infty} i^n J_n(k_0 \rho) e^{in(\theta-\theta_0)}$ (supposing that it is traveling in a direction forming an angle $\theta_0$ with the $x$ axis and that the cylinder axis is along $z$), $\rho$ and $\theta$ being

-16-

respectively the radial and azimuthal coordinates, the magnetic field induced inside the cavity may be written in terms of the same functions, i.e., $\mathbf{H}_c = H_0 \hat{\mathbf{z}} \sum_{n=-\infty}^{\infty} c_n i^n J_n(k_0 \rho) e^{in(\theta-\theta_0)}$, where the coefficients $c_n$ may be found by applying the proper boundary conditions at the two interfaces. A closed form expression for the coefficients may be written by applying Kramer's formula, as follows:

$$c_n = -\frac{\begin{vmatrix} 0 & J_n(k_s a_c) & Y_n(k_s a_c) & 0 \\ 0 & J'_n(k_s a_c)/\varepsilon_s & Y'_n(k_s a_c)/\varepsilon_s & 0 \\ J_n(k_0 a) & J_n(k_s a) & Y_n(k_s a) & H_n(k_0 a) \\ J'_n(k_0 a)/\varepsilon_0 & J'_n(k_s a)/\varepsilon_s & Y'_n(k_s a)/\varepsilon_s & H'_n(k_0 a)/\varepsilon_0 \end{vmatrix}}{\begin{vmatrix} J_n(k_0 a_c) & J_n(k_s a_c) & Y_n(k_s a_c) & 0 \\ J'_n(k_0 a_c)/\varepsilon_0 & J'_n(k_s a_c)/\varepsilon_s & Y'_n(k_s a_c)/\varepsilon_s & 0 \\ 0 & J_n(k_s a) & Y_n(k_s a) & H_n(k_0 a) \\ 0 & J'_n(k_s a)/\varepsilon_s & Y'_n(k_s a)/\varepsilon_s & H'_n(k_0 a)/\varepsilon_0 \end{vmatrix}}, \qquad (8)$$

where $J_n(.)$, $Y_n(.)$ and $H_n(.) = J_n(.) + iY_n(.)$ are the cylindrical Bessel functions of integer order $n$ [28]. The derivatives in the previous formulas are taken with respect to the argument of the Bessel functions.

We note from Eq. (8) that when $a_c = a$ the coefficients simplify to $c_n = 1$, since the ENZ shell would have zero thickness and therefore the field in the hollow region is the same as $\mathbf{H}_i$.

Taking the limit for $\varepsilon_2 \to 0$, expression (8) yields the interesting limit:

$$c_n = \begin{cases} 1 & a_c = a \\ 0 & a_c \neq a \end{cases}, \qquad (9)$$

for any $n \neq 0$, and:

$$c_0 = \frac{i\bar{a}\left[J_0(\bar{a})Y_1(\bar{a}) - Y_0(\bar{a})J_1(\bar{a})\right]/J_0(\bar{a}_c)}{\bar{a}H_1(\bar{a}) - \left(\bar{a}_c \frac{J_1(\bar{a}_c)}{J_0(\bar{a}_c)} + \frac{\bar{a}^2 - \bar{a}_c^2}{2}\right)H_0(\bar{a})}, \qquad (10)$$

being $\bar{a} = k_0 a$ and $\bar{a}_c = k_0 a_c$. This ensures that the ENZ shell with a permittivity sufficiently close to zero would induce an azimuthally constant-phase field inside the cavity, independent of the angle $\theta_0$ from which the plane wave impinges, and more generally, given the linearity of the problem, independent of the form of the excitation. Again, as predicted heuristically, the phase front at the exit side conforms to the shape of the ENZ region, and the ENZ medium "isolates" the two regions that it delimits.

In the limit of $\varepsilon_2 \to 0$, this effect is independent of the thickness of the ENZ shell and independent of the size of the cavity, as Eqs. (9)-(10) show. For finite $\varepsilon_s$, a larger cavity requires a value of $\varepsilon_s$ closer to zero to yield the results predicted by (9)-(10). Figure 9 shows the amplitude of the first three coefficients $c_n$ for a hollow cavity with diameter $2a_c = \lambda_0$, with $\lambda_0$ being the free-space wavelength, varying the shell's outer radius $a$ for two different values of permittivity $\varepsilon_s$. We notice how in the case of Fig. 9a, all the higher order coefficients go to zero very fast, as soon as $a \neq a_c$, as predicted by (9). This implies that, independently of the thickness of the shell, the field inside the cavity does not vary azimuthally, being dominated by the $n=0$ mode. In the case of Fig. 9b, for a larger value of $\varepsilon_s$, you notice how for very thin shells higher-order contributions may become dominant, and the value of the coefficients with $n \neq 0$ goes more slowly to zero than in the previous case with an increase in the shell thickness. For smaller cavities, the required values of $\varepsilon_s$ for getting a dominant $n=0$ term are less close to zero.

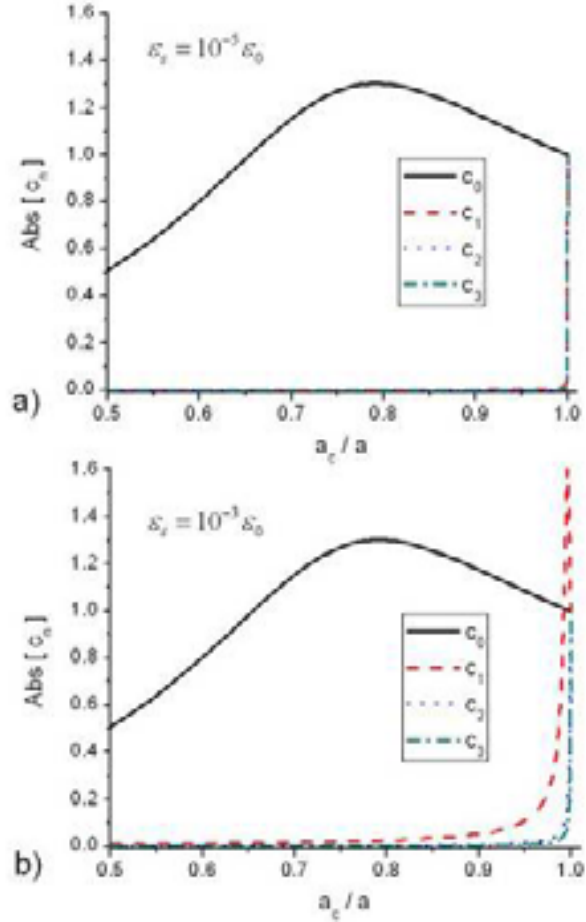

Figure 9 – (Color online) Magnitude of the first three coefficients $c_n$ for a hollow cylinder with $a_c = \lambda_0 / 2$ surrounded by an ENZ shell.

Figure 10 shows the magnitude of the total electric (Fig. 10a) and magnetic (Fig. 10b) fields induced by a TM plane wave traveling along $x$ and impinging on the cylinder of Fig. 7 formed by a cavity with $a_c = \lambda_0$ delimited by an ENZ shell with $a = a_c / 0.8$ and $\varepsilon_s = 10^{-3} \varepsilon_0$. Following the previous analysis, we expect that the relative magnitude of the higher order coefficients is very low when compared with the coefficient $c_0$: in this specific case $|c_1| = 4.54\% |c_0|$, $|c_2| = 0.75\% |c_0|$ and higher order coefficients are negligible. We see how inside the cavity the angular variation of the electric and magnetic fields is not observable, despite the clear asymmetry of the excitation and of the induced fields in the outside region and in the ENZ shell. The field inside the cavity, despite its relatively electrically large size, is essentially a standing wave with distribution $\mathbf{H}_c = H_0 \hat{\mathbf{z}} c_0 J_0(k_0 \rho)$. The modification of the impinging phase pattern due to the ENZ material is very evident in this example, and the isolated cavity here presented may offer interesting potentials for applications.

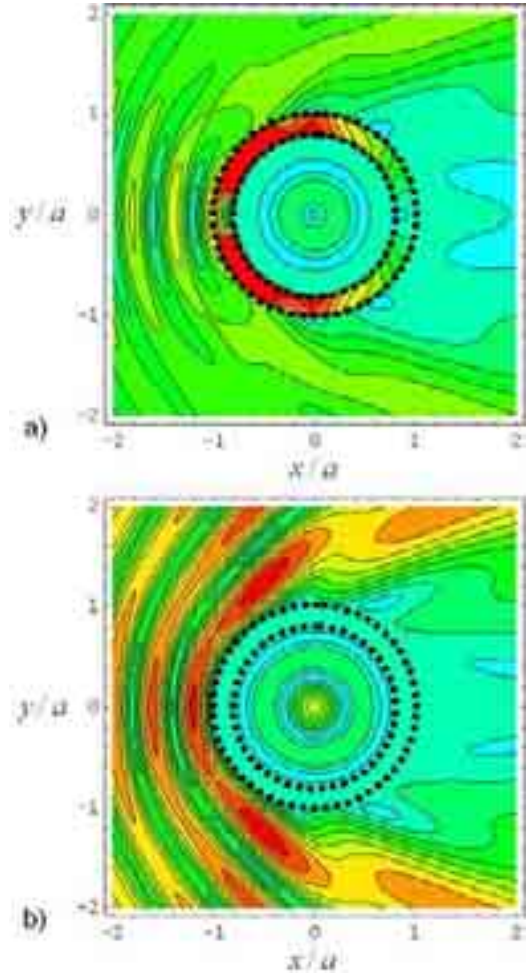

Figure 10 – (Color online) Electric (a) and magnetic (b) field distributions for a TM plane wave traveling along $x$ and impinging on the cylindrical ENZ shell of Fig. 7 with $a_c = \lambda_0$, $a = a_c/0.8$, $\varepsilon_s = 10^{-3}\varepsilon_0$. Brighter (more red) colors correspond to higher amplitudes of the field.

Figure 11 shows the power flow (i.e., the real part of the Poynting vector) for the same geometry as in Fig. 10. In this case, similar to the planar geometry, the ENZ shell redirects the power flow in order to isolate the inner cavity from the asymmetric phase pattern of the source excitation. Since the cavity in this case is electrically large (its diameter is two wavelengths) some minor circulation of power is visible, due to the interaction with higher-order modes inside the cavity, whereas for smaller cavities the inner region is essentially totally isolated without any local net power flow. The inner power flow feeds the resonant circulation of power arising in the shell region and keeps established the cavity mode that we have predicted in the previous lines. This circulation and re-direction of power provided by the ENZ shell reminds also of the similar power flow distribution induced in the phenomenon of scattering suppression and induced invisibility that we have obtained employing such ENZ shells in a different configuration [22].

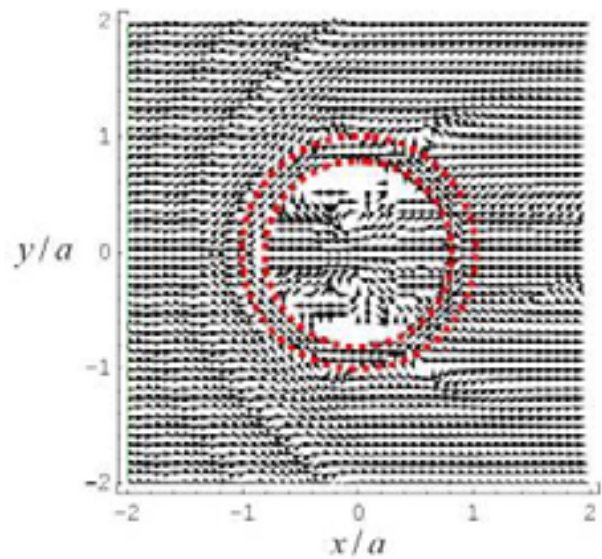

Figure 11 – (Color online) Power flow (real part of the Poynting vector) distribution for the geometry of Fig. 9.

The presence of material losses does not affect strongly these results, since these effects do not rely on strong polaritonic resonances of the metamaterial shell. The suppression effect of the higher-order modes is weakened by the presence of material losses, but realistic materials (i.e., with a loss tangent of about $10^{-2}$) may still support these findings.

As in the planar case, the TE polarization does not respond similarly to the presence of the ENZ shell. This is consistent with the previous considerations.

## 5. *Method of Moments simulations with more complex geometries*

In order to confirm the results reported in the previous sections and to demonstrate the generality of the concepts here discussed and show novel possibilities, here we study numerically the scattering of waves from complex shaped ENZ obstacles. The following numerical simulations have been obtained with a home-made Method-of-Moments (MoM) simulator for 2D cylindrical structures.

In the first example, we validate the results of the previous sections, investigating whether an ENZ slab with $\varepsilon_s = 0.01$ may behave as a lens redirecting the incoming energy towards a specific direction in space. To this end, the output face of the object is tilted with respect to the input face, as illustrated in Fig. 12. An incoming plane wave (propagating along the positive *x*-direction) illuminates this 2-D trapezoidal shaped obstacle. Consistently with the heuristic and theoretical results of the previous sections, it is seen in Fig. 12 how the phase of the transmitted field is nearly parallel to the output face of the ENZ slab. This phenomenon is to some extent independent of the characteristics of the field impinging on the entrance face. This is illustrated in Fig. 13, where we plot the phase of the transmitted magnetic field when a magnetic line source is scanned along the direction parallel to the input face of the ENZ slab. It is seen that even though this singular source is very close to the ENZ object, the wavefronts of the transmitted wave remain to some extent parallel to the exit face. In other words, the ENZ trapezoidal cylinder under analysis and similar geometries may act as lenses, basically focusing the energy from a source and reshaping the phase pattern. Compared to standard lenses, however, this configuration is not limited by restrictions in size, being possibly very thin and conformal to the desired shape. Also this and the previous results show how the focal surfaces may be shaped according to the necessity, providing larger degrees of freedom in their design and in the possibility of employment with respect to common lenses. As the permittivity of the object approaches zero, the wavefronts become increasingly conformal to the exit face.

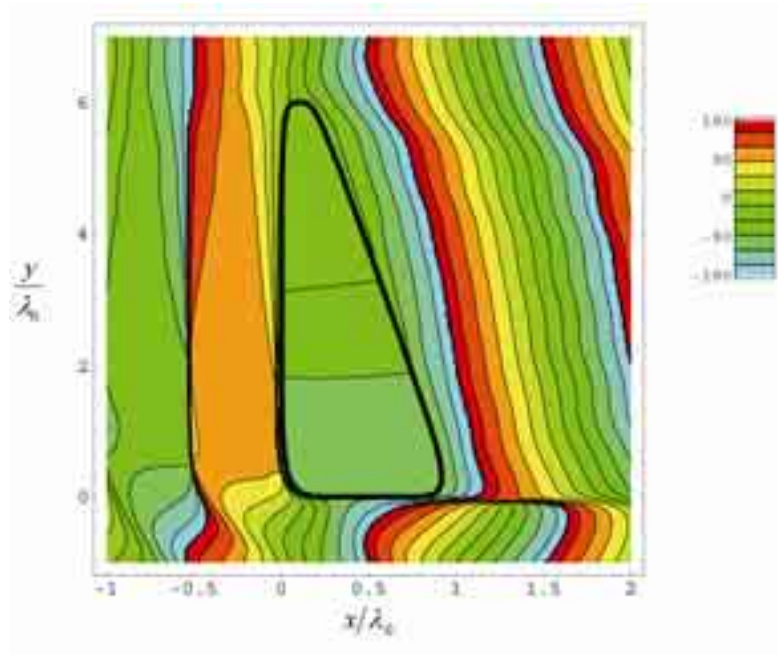

Figure 12 - (Color online) Phase of the magnetic field distribution (in degrees) when a TM plane wave impinges on a 2-D trapezoidal slab with $\varepsilon_s = 0.01\varepsilon_0$.

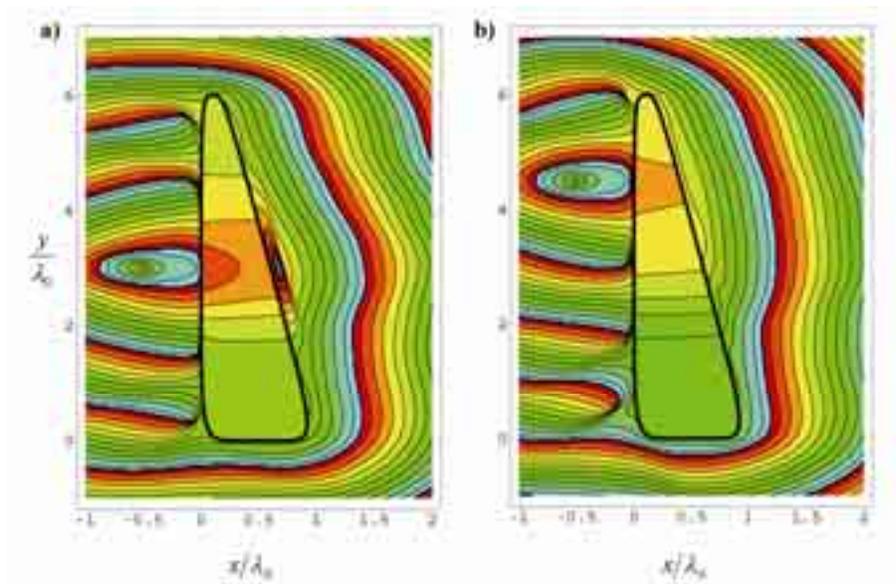

Figure 13 - (Color online) Phase of the magnetic field distribution when a magnetic line source illuminates the 2-D trapezoidal scatterer of Fig. 12. a) line source positioned at $(-0.5, 3.0)\lambda_0$. b) line source positioned at $(-0.5, 4.5)\lambda_0$. The color scale is as in Fig. 12.

This remarkable property of ENZ objects is indeed completely general, and not specific of objects with planar interfaces or other canonical shapes. Indeed, as shown in Fig. 14, this property is revealed even for objects with irregular or "randomly" shaped interfaces.

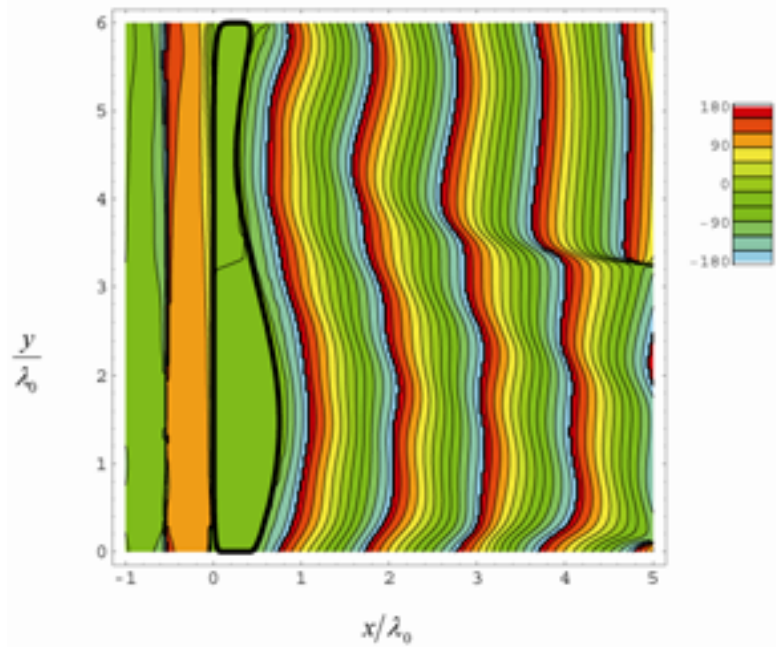

Figure 14 - (Color online) Phase of the magnetic field distribution when a plane wave illuminates a 2-D cylindrical object with $\varepsilon_s = 0.01\varepsilon_0$ and arbitrarily shaped output interface.

The previous results suggest the possibility of using a concave ENZ "lens" to transform an incoming planar wavefront into a convergent cylindrical wavefront, consistent to what reported in the previous sections and with the geometry of Fig. 7. To investigate such possibility we have analyzed the field transmitted by a 2-D ENZ lens with planar input interface, and concave output interface with radius of curvature $6\lambda_0$. In Fig. 15 the amplitude (Fig. 15a) and phase (Fig. 15b) of the magnetic field are plotted assuming that the exciting field is a plane wave impinging normal to the entrance face of the slab. Consistently with our theory, it is seen that the electromagnetic field is focused at the center of curvature of the output interface, around $(3.0, 6.2)\lambda_0$. We have also verified that the focal spot tends to be narrower if the permittivity of the slab is brought to a value closer to zero. In addition, we have verified that the same effect occurs when the slab is illuminated by a magnetic line source positioned at different locations, as reported in Fig. 16. For this specific configuration, we have assumed also the permeability of the lens being less than the one of free space, i.e., $\varepsilon_s/\varepsilon_0 = \mu_s/\mu_0 = 0.01$, in order to have a matched lens and higher transmission from the structure.

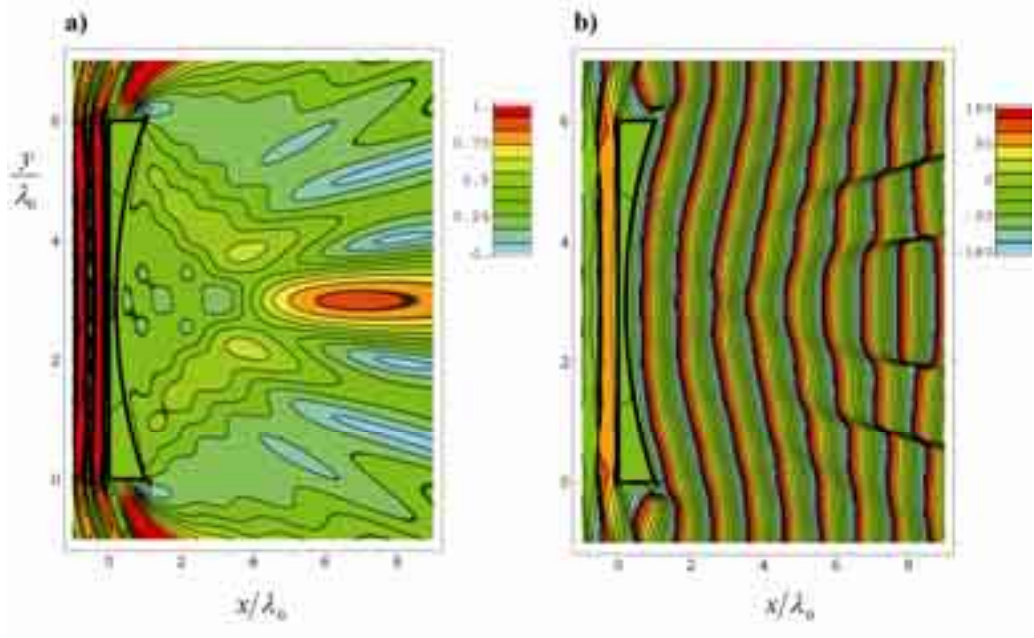

Figure 15 - (Color online) Normalized amplitude (a) and phase (b) of the magnetic field distribution when a plane wave illuminates a 2-D concave lens with $\varepsilon_s = 0.01\varepsilon_0$ and radius of curvature $6\lambda_0$.

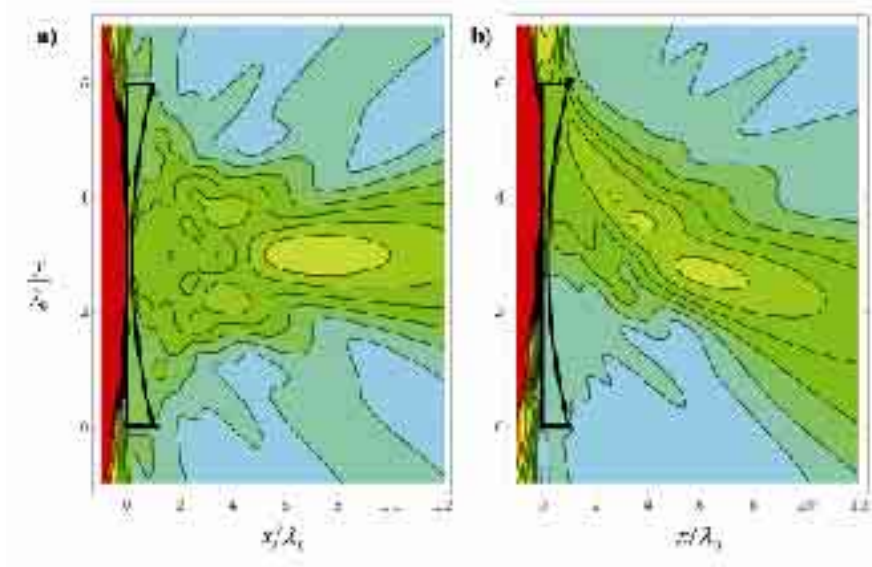

Figure 16 - (Color online) Normalized amplitude of the magnetic field distribution when a line source illuminates the concave lens of Fig. 15 with $\varepsilon_s/\varepsilon_0 = \mu_s/\mu_0 = 0.01$ and radius of curvature $6\lambda_0$. a) line source positioned at $(-0.5, 3.0)\lambda_0$. b) line source positioned at $(-0.5, 4.5)\lambda_0$. The color scale is as in panel a) of Fig. 15.

Conversely, if the output face of the 2-D ENZ object is convex, such plasmonic lens may transform an incident arbitrary wave into a divergent cylindrical wave. This interesting possibility is supported by the results shown in Fig. 17, which corresponds to the case in which the convex object is illuminated by a magnetic line source that is scanned along the planar input interface. The radius of curvature of the convex interface is $6\lambda_0$. As in the previous cases, also here the location of the line source does not influence the phase pattern in the output region.

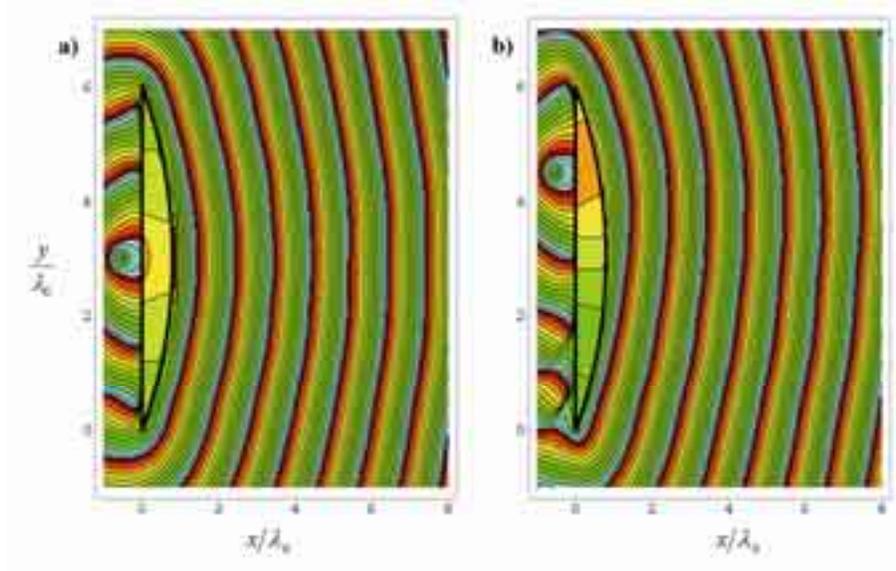

Figure 17 - (Color online) Phase of the magnetic field distribution when a point source illuminates an ENZ convex lens with $\varepsilon_s = 0.01\varepsilon_0$ and radius of curvature $6\lambda_0$. a) line source positioned at $(-0.5, 3.0)\lambda_0$. b) line source positioned at $(-0.5, 4.5)\lambda_0$. The color scale is as in Fig. 12.

Since we have shown how an ENZ lens may effectively control the phase front of the transmitted wave with the shape of its exit face, it is interesting to study what would happen if two or more of these objects are combined in series so that the incoming wave undergoes successive phase transformations. In order to ensure relatively good coupling between adjacent ENZ objects, it is clear from the previous discussion how the output interface of the first ENZ object should be complementary to the input interface of the adjacent ENZ object. In Fig. 18 we have studied the geometry where two 2-D complementary trapezoidal shaped ENZ lenses are paired together and illuminated by a plane wave propagating along the positive *x*-direction. It is seen that the wavefronts become conformal to the shape of the first object after the first transmission, and that, after transmission through the second object, become again parallel to the incoming wave. In Fig. 19 we show the phase of the magnetic field distribution for the scenario in which the input interface of the first object and the output interface of the second object are not parallel. In this situation, the wavefronts (after two successive transmissions) are tilted with respect to the incoming wave. As illustrated in Fig. 20, similar results are obtained if concave/convex complementary objects are paired together.

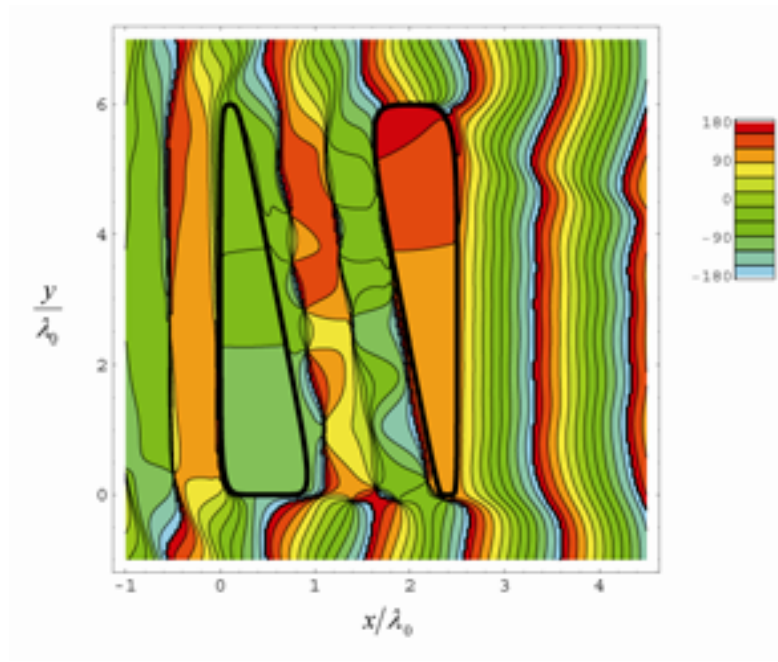

Figure 18 - (Color online) Phase of the magnetic field distribution when a plane wave (propagating along the positive *x*-direction) illuminates two complementary ENZ lenses with $\varepsilon_s = 0.01\varepsilon_0$.

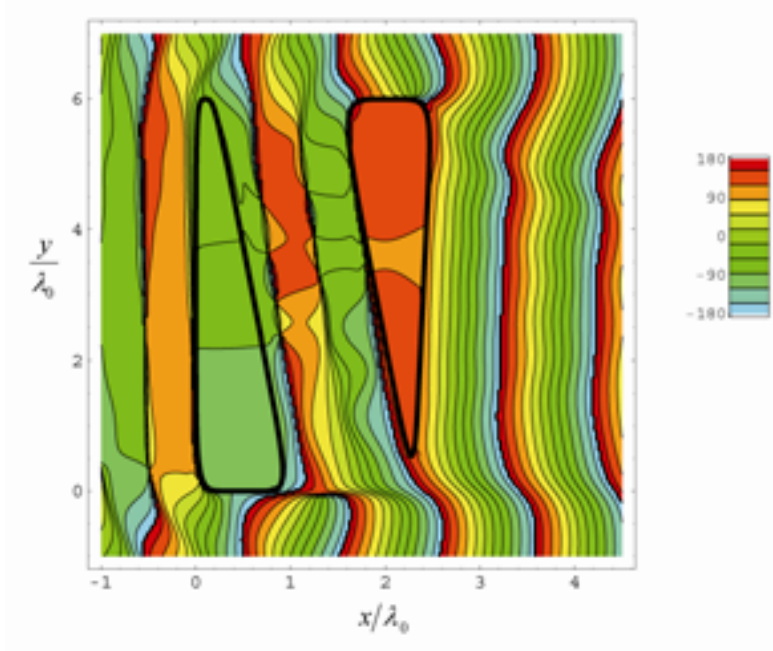

Figure 19 - (Color online) Phase of the magnetic field distribution when a plane wave (propagating along the positive $x$-direction) illuminates two complementary objects with $\varepsilon_s = 0.01\varepsilon_0$. The output face of the second object is tilted with respect to the input face of the first object, causing a tilting of the transmitted wavefronts.

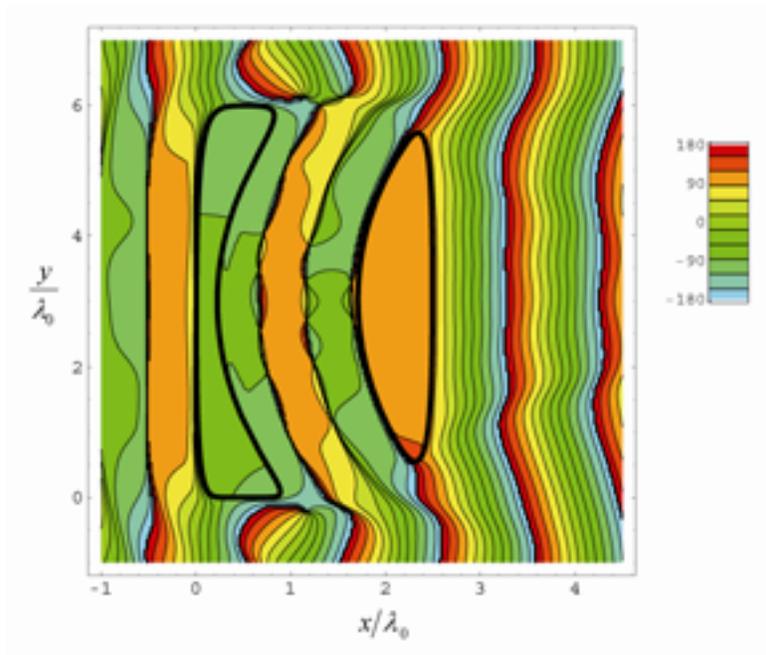

Figure 20 - (Color online) Phase of the magnetic field distribution when a plane wave (propagating along the positive $x$-direction) illuminates two complementary ENZ lenses with $\varepsilon_s = 0.01\varepsilon_0$. The first object has a concave exit face and the second object has a convex entrance face.

As shown analytically in the previous section, another interesting application of the phase-restoration properties of ENZ materials is represented by the possibility of shielding a cavity or a hole from the phase variations induced by an external source. In this way it is ensured that the angular phase variation of the field inside a circular cavity is nearly uniform. Here we compute the field distribution for a more complex setup with respect to the previous section, i.e., for a cavity embedded in a planar finite slab of ENZ material.

In Fig. 21 it is reported the case of a plane wave illuminating an ENZ slab with a small circular hollow cavity with radius $R = 0.125\lambda_0$. It is visible how both amplitude and phase of the magnetic field distribution are nearly uniform inside the ENZ slab and inside the hollow cavity, even though the geometry is asymmetrical.

A related geometry that has also attracted our attention is simulated in Fig. 22. It consists of an ENZ slab in which a small semi-circular cavity with radius $R = 0.3\lambda_0$ was carved in the exit face. We tested if, when illuminated by a plane wave, the phase front at the output face would be tailored by the shape of the small cavity, even though the radius of the cavity is electrically small. As it is apparent in Fig. 22a, indeed the phase fronts conform to the anomalous surface shape. In Fig. 22b we have studied the response of such structure to a line source placed at the center of such hemicylinder. In this case, the radiated beam is collimated into the direction normal to the planar interface.

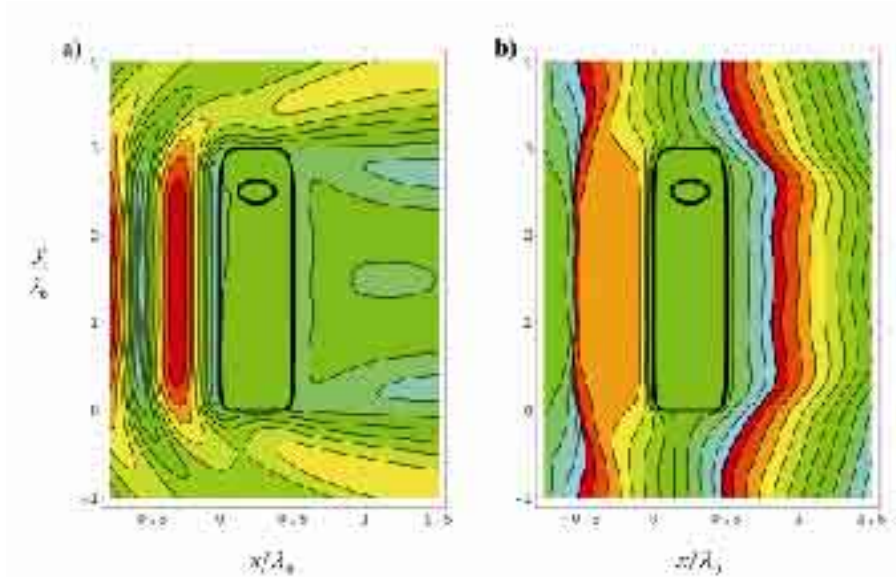

Figure 21 - (Color online) Normalized amplitude (a) and phase (b) of the magnetic field distribution when a plane wave propagating along the positive *x*-direction illuminates an ENZ slab with $\varepsilon_s = 0.01\varepsilon_0$ with a hollow cylindrical cavity of permittivity $\varepsilon_0$ and radius $R = 0.125\lambda_0$. The color scale is as in Fig. 15.

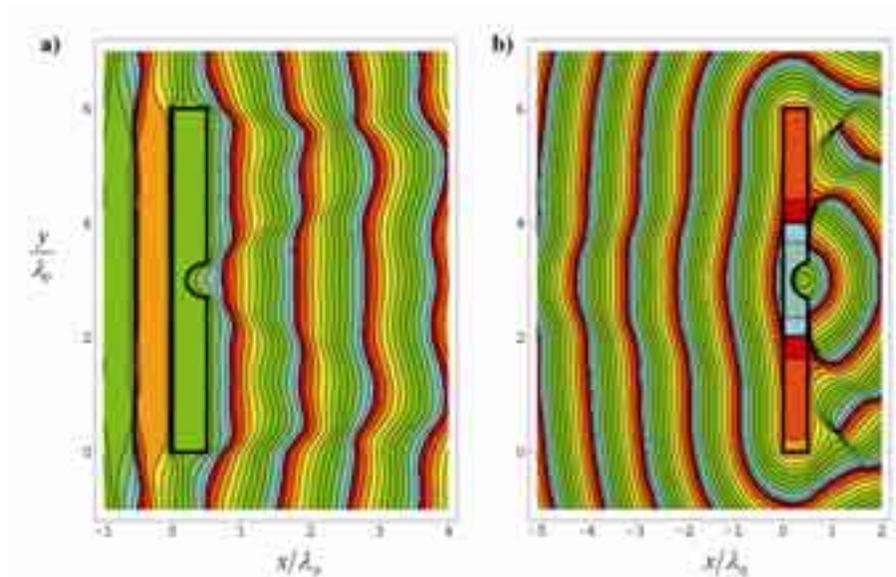

Figure 22 - (Color online) Phase of the magnetic field distribution when: (a) a plane wave impinges from the negative *x*-direction; (b) a point source is positioned at $(0.5, 3.0)\lambda_0$, in both cases illuminating a 2-D ENZ planar slab with $\varepsilon_s = 0.01\varepsilon_0$ with a small hemi-cylindrical cavity carved on its exit face with radius $R = 0.3\lambda_0$. The color scale is as in Fig. 12.

## 6. Engineering the phase pattern

It may be of interest for practical purposes to analyze the inverse problem of determining the geometry of the ENZ lens required to transform a given incident wavefront $\psi^{inc}$ into another desired wavefront. Our solution in this section is based on geometrical optics considerations [29]. Within the validity of this approximation one can assume that the electromagnetic fields are modulated by the propagation factor $e^{ik_0\psi}$, where $k_0$ is the free-space wave number and $\psi$ is the eikonal. As well-known [29], the eikonal is the solution of the first order partial differential equation $|\nabla \psi| = n$, where $n$ is the index of refraction of the medium.

Suppose now that $\psi$ is prescribed at two different planes $x = x_0$ and $x = x_f$ (apart from the sum of an irrelevant phase constant). The problem is to determine the lens shape that guarantees such phasefront transformation. Assuming that the lens material has index of refraction close to zero, the input and output faces of the lens can be designed nearly independently, because they are coincident with the wavefronts in free-space.

So the problem of determination of the lens shape is equivalent to calculating the desired wavefront shapes in free-space and then filling the space in the middle with an ENZ material acting as the desired lens.

Suppose that at the output plane $x = x_f$, the prescribed phase is given by:

$$\psi|_{x=x_f} = f(y) \tag{11}$$

Since the index of refraction of free-space is unity, the eikonal equation becomes:

$$\nabla \psi|_{x=x_f} = \sqrt{1-\dot{f}^2}\,\hat{\mathbf{u}}_x + \dot{f}\,\hat{\mathbf{u}}_y \tag{12}$$

where $\dot{f} = df/dy$.

The eikonal equation may be solved using the Hamilton method, which involves calculating the characteristic curves of the system, i.e. the rays [29]. In homogeneous media, the rays are tangent to $\nabla \psi$ and the eikonal varies along a ray as $\psi = \psi_0 + ns$ (with $n = 1$ in free-space and $s$ being the distance traveled along the ray). Consequently, at the output region the wavefronts are defined by the following family of parallel surfaces, parameterized relatively to $y$:

$$S = (x_f, y) + \frac{\nabla \psi|_{x=x_f} (C - \psi(y))}{n} \tag{13}$$

where $C$ is an arbitrary constant. There is a one-to-one correspondence between $C$ and the wavefronts, i.e. by fixing $C$ a specific wavefront at the output region is selected.

Thus, the previous analysis demonstrates that the desired ENZ lens may be systematically designed by: (a) calculating the wavefronts associated with the input and output planes, as described above; (b) selecting two specific wavefronts, one associated with the input face and the other with the output face (this selection is in general arbitrary, apart from the fact that the two wave fronts should not be too far apart, since the ENG lens should be desirably thin to ensure reasonable transmission); (c) filling the region of space delimited by the two wavefronts with an ENZ material.

To illustrate these concepts we consider here as an example the case in which the incoming wavefront is planar, and consequently the input face of the lens is also designed to be planar. The aim of the design is to shape the outer interface so that the prescribed eikonal is $\psi = 0.05(y - 3.0)^2$ at $x = x_f = \lambda_0$ over $0 < y < 6\lambda_0$. Following the outlined algorithm, one obtains the ENZ lens shape depicted in Fig. 23. In the figure it is reported a MoM simulation for the phase distribution in the free space region. Fig. 24, moreover, compares the eikonal $\psi$ at the output face compared with the desired target function. For filling the lens region we considered $\varepsilon_s = 0.01\varepsilon_0$ and TM polarization in the simulation. It is seen that a good agreement is obtained apart from a small error at the edges of the lens.

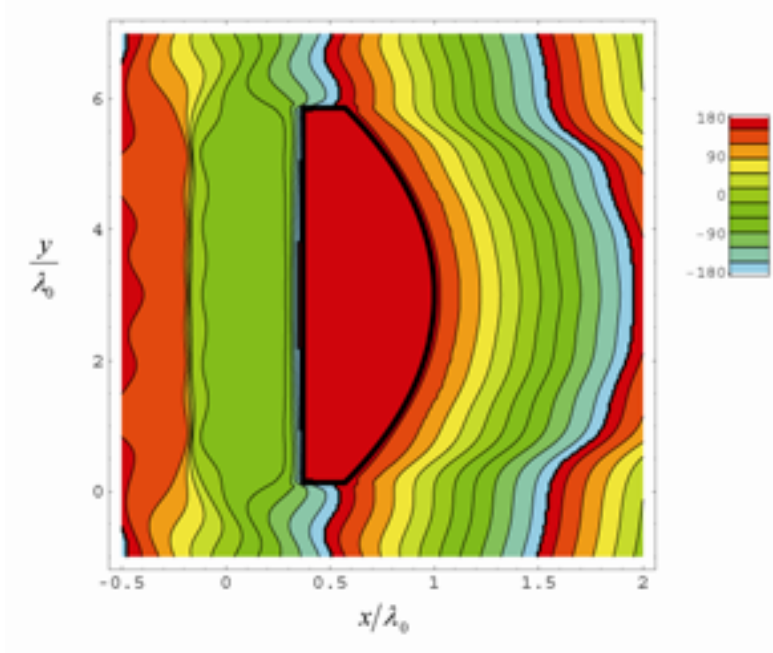

Figure 23 - (Color online) Phase of the magnetic field distribution when a plane wave (propagating along the positive x-direction) illuminates a 2-D lens with $\varepsilon_s = 0.01\varepsilon_0$ designed to obtain a prescribed phase front at the output plane.

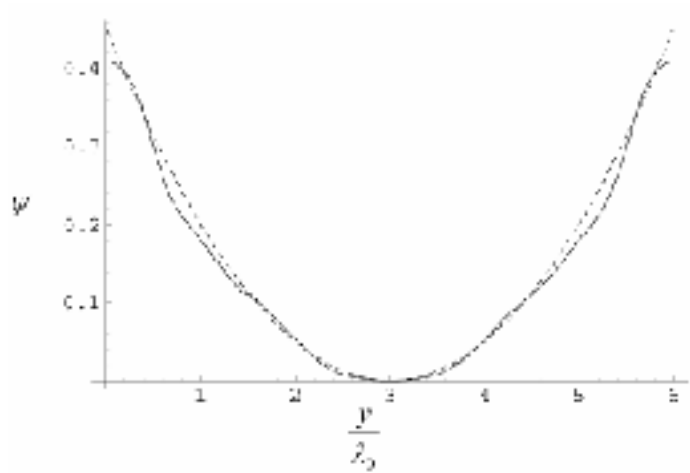

Figure 24 - Eikonal $\psi$ as a function of $y$ at the exit face of the ENZ lens of Fig. 23, i.e., at $x = x_f = \lambda_0$. Solid line: synthesized wavefront. Dashed line: target wavefront.

## 7. *Conclusions*

In this work, we have fully analyzed the response of ENZ materials to electromagnetic excitation. In particular, we have studied the possibility of using these materials in various geometries to tailor the phase pattern of the output radiation when the structure is excited by an arbitrary source. Planar slabs, cylindrical shells, complementary double ENZ structures, and ENZ lenses with various exit faces have been shown to effectively isolate the regions of space they delimit and tailor the phase pattern according to the shape of their interfaces. An algorithm to synthesize lenses tailoring prescribed arbitrary wavefronts has also been presented. The described effects, although dependent on the polarization of the field, may have interesting potential applications in several fields.

*Acknowledgements*

This work is supported in part by the U.S. Air Force Office of Scientific Research (AFOSR) grant number FA9550-05-1-0442. Andrea Alù was partially supported by the 2004 SUMMA Graduate Fellowship in Advanced Electromagnetics. Mário Silveirinha has been partially supported by a grant from "Fundação para a Ciência e Tecnologia".

*References*

[*]   To whom correspondence should be addressed, engheta@ee.upenn.edu

[1]   J. B. Pendry, Phys. Rev. Lett. **85**, 3966 (2000).

[2]   R. W. Ziolkowski and E. Heyman, Phys. Rev. E, **64**, 056625 (2001).

[3]   D. R. Smith, D. Schurig, and J. B. Pendry, Appl. Phys. Lett. **81**, 2713 (2002).

[4]   V. M. Shalaev, W. Cai, U.K. Chettiar, H.-K. Yuan, A.K. Sarychev, V.P. Drachev, and A.V. Kildishev, Opt. Lett. **30**, 3356 (2005).

[5]   G. Dolling, C. Enkrich, M. Wegener, J. F. Zhou, C. M. Soukoulis, and S. Linden, Opt. Lett. **30**, 3198 (2005).

[6]   S. Zhang, W. Fan, N. C. Panoiu, K. J. Malloy, R. M. Osgood, and S. R. J. Brueck, Phys. Rev. Lett. **95**, 137404 (2005).

[7]   A. N. Grigorenko, A. K. Geim, H. F. Gleeson, Y. Zhang, A. A. Firsov, I. Y. Khrushchev, and J. Petrovic, Nature **438**, 335 (2005).

[8]  C. Bohren and D. Huffmann, *Absorption and Scattering of Light by Small Particles* (John Wiley, New York, 1983).

[9]  J. D. Jackson, *Classical electrodynamics*, (John Wiley, New York, 1999).

[10] J. Gómez Rivas, C. Janke, P. Bolivar, and H. Kurz, Opt. Express **13**, 847 (2005).

[11] P. B. Johnson and R. W. Christy, Phys. Rev. B **6**, 4370 (1972).

[12] J. Brown, Proc. IEE **100 IV**, 51 (1953).

[13] I. J. Bahl and K. C. Gupta, IEEE Trans. Ant. Prop. **22**, 119 (1974).

[14] S. A. Kyriandou, R. E. Diaz, and N. G. Alexopoulos, Dig. of 1998 IEEE Antennas Propagat. Int. Symp., Atlanta, Georgia, June 21-26, 1998, pp.660-663, Vol. 2.

[15] S. Enoch, G. Tayeb, P. Sabornoux, N. Guerin, and P. Vincent, Phys. Rev. Lett., **89**, 213902 (2002).

[16] J. Pacheco, T. Gregorczyk, B. I. Wu, and J. A. Kong, PIERS 2003, Honolulu, Hawaii, Oct. 13-16, 2003, p. 479.

[17] N. Garcia, E. V. Ponizovskaya, and John Q. Xiao, Appl. Phys. Lett. **80**, 1120 (2002).

[18] R. W. Ziolkowski, Phys. Rev. E **70**, 046608 (2004).

[19] A. Alù, F. Bilotti, N. Engheta, and L. Vegni, IEEE Trans. Antennas Propagat. **54**, 1632 (2006).

[20] P. Baccarelli, P. Burghignoli, F. Frezza, A. Galli, P. Lampariello, G. Lovat, and S. Paulotto, IEEE Trans. Microwave Th. Tech. **53**, 32 (2005).

[21] M. Silveirinha and N. Engheta "Squeezing Electromagnetic Energy through Sub-wavelength Channels and Bends Using ENZ Materials", Physical Review Letters (to be published, 2006).

[22] A. Alù, and N. Engheta, Phys. Rev. E, **72**, 016623 (2005).

[23] R. Englman, and R. Ruppin, J. Phys. C, **1**, 614 (1968).

[24] R. Ruppin, and R. Englman, J. Phys. C, **1**, 630 (1968).

[25] R. Englman, and R. Ruppin, J. Phys. C **1**, 1515 (1968).

[26]     CST Microwave StudioTM 5.0, CST of America, Inc., www.cst.com.

[27]     Comsol Multiphysics 3.2, COMSOL inc. www.comsol.com.

[28]     M. Abramowitz and I. Stegun, *Handbook of Mathematical Functions* (Dover Publications, Inc., New York, 1970).

[29]     M. Born and E. Wolf, *Principles of Optics* (Cambridge University Press, 1999).